\documentclass[aps,pra,twocolumn,10pt]{revtex4-1}

\usepackage[english]{babel}
\usepackage{amsmath}
\usepackage{amsfonts}
\usepackage{amssymb}
\usepackage{graphicx}
\usepackage{siunitx}
\usepackage{hyperref}
\usepackage[english,capitalise]{cleveref}

\renewcommand{\i}{\mathrm{i}}
\renewcommand{\vec}[1]{\boldsymbol{#1}}
\renewcommand{\Im}{\mathop{\mathrm{Im}}}

\begin{document}

\title{Population transfer at exceptional points in spectra of the hydrogen atom in parallel electric and magnetic fields}

\author{Lukas Oberreiter}
 \email{lukas.oberreiter@itp1.uni-stuttgart.de}
\author{Jan Burkhardt}
\author{J{\"o}rg Main}
\author{G{\"u}nter Wunner}

\date{\today}

\affiliation{Institut f{\"u}r Theoretische Physik 1, Universit{\"a}t Stuttgart, 70550 Stuttgart, Germany}

\begin{abstract}
We study the population transfer between resonance states for a
time-dependent loop around exceptional points in spectra of the
hydrogen atom in parallel electric and magnetic fields. Exceptional
points are well-suited for population transfer mechanisms, since a
closed loop around these in parameter space commutes eigenstates.
We address the question how shape and duration of the dynamical
parameter loop affects the transferred population, in order to
optimize the latter. Since the full quantum dynamics of the expansion
coefficients is time-consuming, we furthermore present an
approximation method, based on a $2\times 2$ matrix.
\end{abstract}


\maketitle

\section{Introduction}
\label{sec:introduction}
Open quantum systems can effectively be described by
\emph{non-hermitian} Hamiltonians. With the method of complex scaling
originally hermitian operators can be transformed to the non-hermitian
domain \cite{reinhardt_complex_1982,moiseyev_quantum_1998}. The
advantage of the non-hermitian description is that expensive
time-dependent calculations are avoided, but instead the
\emph{stationary} Schr{\"o}dinger equation which yields complex
eigenvalues is solved \cite{moiseyev_non-hermitian_2011}. An example
is the case of resonances, which are discrete metastable states living
in the lower half of the complex energy plane. The bound states of the hydrogen
atom become resonances by applying a constant electric field, which
deforms the potential such that the electron can tunnel through a
barrier to the continuum.

In spectra of non-hermitian Hamiltonians exists a special kind of degeneracy, called \emph{exceptional point} (EP), which is a point in parameter space, where (at least) two eigenstates coalesce.
To bring two eigenstates to coalescence an (at least) two-dimensional real parameter space is necessary.

As is well known a  simple example \cite{kato_perturbation_1976} to demonstrate properties of EPs is the non-hermitian matrix
\begin{equation}
\mathcal{M}(\kappa)=
\begin{pmatrix}
1 &  \kappa \\
\kappa  & -1 
\end{pmatrix}
\quad \mathrm{with}\ \kappa \in \mathbb{C} \ ,
\end{equation}
for which the right eigenvectors $\vec{v}_i$ and eigenvalues $\lambda_i$ read
\begin{equation}
\vec{v}_{1,2}(\kappa) =
\begin{pmatrix}
-\kappa \\
1\mp \sqrt{1+\kappa^2} 
\end{pmatrix}
,\ \  
\lambda_{1,2} = \pm \sqrt{1+\kappa^2}
\ .
\end{equation}
EPs exist for the parameter values $\kappa=\pm \i$.
The eigenvalue surface is divided into two Riemann sheets, which both possess a single value at the EP. A single loop around the EP in the complex parameter space $\kappa$ commutes the two eigenvalues, whereas two loops would rearrange the initial configuration.

Examples for theoretical treatments of EPs in quantum systems are atomic \cite{latinne_laser-induced_1995,magunov_strong_1999,magunov_laser-induced_2001,holger_cartarius_exceptional_2007} and molecular \cite{lefebvre_resonance_2009,Leclerc2017} spectra, scattering of particles at potential barriers \cite{korsch_stark_2003,hernandez_non-hermitian_2006}, atom waves \cite{holger_cartarius_discovery_2008,rapedius_nonlinear_2010,gutohrlein_bifurcations_2013,abt_supersymmetric_2015}, open Bose-Hubbard systems \cite{graefe_non-hermitian_2008}, unstable lasers \cite{v._berry_mode_2003}, resonators \cite{klaiman_visualization_2008}, and optical waveguides \cite{wiersig_asymmetric_2008,wiersig_enhancing_2014}. 
Furthermore there exists experimental evidence of EPs. They have been shown to exist in metamaterials \cite{lawrence_manifestation_2014}, a photonic crystal slab \cite{zhen_spawning_2015}, electronic circuits \cite{stehmann_observation_2004}, microwave cavities \cite{philipp_frequency_2000,dembowski_observation_2003,dietz_rabi_2007,chen_exceptional_2017},
microwave waveguides \cite{Doppler2016}, and
exciton-polariton resonances \cite{gao_observation_2015}.

The system we study is the hydrogen atom in parallel electric and magnetic fields, exhibiting resonances within the non-hermitian description.
Crossed external fields have been used in former studies on the hydrogen atom \cite{henri_menke_state_2016, holger_cartarius_exceptional_2007}. The choice of parallel fields imposes a higher symmetry on the Hamiltonian, which simplifies the calculations and EPs still occur in the simplified system. 
With two field strengths a two-dimensional control space is at hand and EPs can be found. 
If an EP is dynamically encircled the above-mentioned commutation
behavior of two states can be used to transfer population between
resonances. Therefore, at a parameter value close to an EP, the system
population is first prepared, via laser excitation, in one of the resonance states, which would become degenerate directly at the EP. Then the system is perturbed such that the time-dependent parameters form a closed loop around the EP. During the loop the initially populated resonance commutes the position with another resonance state, and if some population remains in the ``wandering'' resonance, population has been transferred to the initially unpopulated state after the loop has been finished.

Recently, in the adiabatic regime, i.e. for slowly time-dependent loops, intensive research has been carried out as regards this kind of population transfer mechanism involving EPs \cite{uzdin_observability_2011,henri_menke_state_2016,berry_slow_2011,gilary_asymmetric_2012,gilary_time-asymmetric_2013-1,graefe_breakdown_2013,viennot_almost_2014,leclerc_discussion_2013}. It has been shown that only the most stable resonance satisfies the adiabatic approximation. All other resonance populations end up in the most stable one.
This is because the adiabatic theorem is not valid in general for non-hermitian Hamiltonians---due to its complex-valued eigenvalues 
\cite{nenciu_adiabatic_1992,dridi_adiabatic_2010}.
It is only valid for tracing the most stable resonance. In other cases a reduction of the parameter variation per time unit cannot compensate the increasing non-adiabatic couplings.
One might wonder whether EPs are even relevant for population transfer mechanisms, since in the long-time limit all population drifts into the long-living resonance. However, if two similar parameter circles are compared, (i) encircles an EP, and (ii) does not encircle the EP, than the transfer in (i) is greater by orders of magnitude than in (ii) \cite{gilary_time-asymmetric_2013-1,henri_menke_state_2016}.
The adiabatic limit can be calculated by integrating over the time-dependent eigenvalue trajectory. In several experiments complex eigenvalue trajectories have been measured and the corresponding adiabatic limit calculated \cite{gao_observation_2015-1,dembowski_experimental_2001,dietz_exceptional_2011,bittner_scattering_2014}.

Here we discuss the question how the \emph{absolute} population transfer from an initially totally occupied resonance to an initially empty resonance state can be optimized by variation of the dynamical parameter loop around an EP.
It is important to note, that out aim differs from the work of Ref.~\cite{henri_menke_state_2016}, where weighted coefficients $\bar a_i$ have been introduced in there Eq.~(7), which completely ignore the decay of the resonances. Figure~2 in Ref.~\cite{henri_menke_state_2016} shows that for an asymmetric state flip the absolute population can easily decrease by about ten orders of magnitude.
Formally the population transfer is a functional of the parameter loop trajectory, and hence it is an infinite-dimensional optimization problem. Furthermore we construct a $2\times 2$ matrix model, from which an approximate population dynamics of the two EP resonances can be calculated. 

The paper is organized as follows. In Sec.~\ref{sec:hatom_in_parallel_fields} we introduce the system under consideration, viz.\ the hydrogen atom in parallel electric and magnetic fields, and present accurate methods to solve the stationary and time-dependent Schr{\"o}dinger equations. We then present an approximate two-dimensional matrix model. Dynamical encirclings of exceptional points are treated in Sec.~\ref{sec:dyn_encircl}. The aim of the section is to optimize the population transfer by variation of the closed parameter loop. In Sec.~\ref{sec:conclusion} we summarize and draw conclusions.

\section{Hydrogen atom in parallel electric and magnetic fields}
\label{sec:hatom_in_parallel_fields}

\subsection{Computation of resonances by complex scaling}

Throughout we use atomic Hartree units, i.e.\ $\hbar=4\pi\epsilon_0=e=m_e=1$. The parallel electric and magnetic fields are assumed to point in $z$-direction:
\begin{equation}
\vec{F} = \frac{F}{F_0}\vec{e}_z\equiv f\vec{e}_z \ ,\quad \vec{B} =\frac{B}{B_0}\vec{e}_z\equiv \gamma\vec{e}_z
\ ,
\end{equation}
with $F_0=\SI{5.14E11}{V/m}$ and $B_0=\SI{2.35E5}{T}$.
Via the principle of minimal coupling these fields enter the stationary Schr{\"o}dinger equation of the unperturbed hydrogen atom, where relativistic effects and the finite nuclear mass are neglected \cite{wunner_atoms_1994}:
\begin{equation}
\mathcal{H}\psi = \left(
-\frac{1}{2}\Delta - \frac{1}{r} + \frac{\gamma\mathrm{L}_z}{2} + \frac{\gamma^2}{8}\left( x^2+y^2 \right) + fz
\right)\psi
= E\psi
\ ,
\label{eq:stat_SG_1}
\end{equation}
where $L_z$ is the $z$ component of the angular momentum operator, and $r=\sqrt{x^2+y^2+z^2}$ the radial distance from the electron to the core. 
The stationary solution of \cref{eq:stat_SG_1} is expanded in terms of
Coulomb-Sturmian functions $\phi_{nlm}(\vec{r})$ \cite{caprio_coulomb-sturmian_2012,zamastil_calculation_2007}:
\begin{equation}
\psi(\vec{r}) = \sum_{nlm} v_{nlm} \phi_{nlm}(\vec{r})
\ .
\label{eq:cs_ansatz}
\end{equation}
The Coulomb-Sturmian functions are radial scaled hydrogen wave functions.
Note that throughout this paper $n$ is reserved for the \emph{radial} quantum number, whereas the principal quantum number is denoted by $N$.
To calculate resonance states the complex scaling method $r\rightarrow b r$, where $b\in\mathbb{C}$, has to be applied to the wave function and Hamiltonian \cite{reinhardt_complex_1982,moiseyev_quantum_1998}.
To obtain a matrix representation of \cref{eq:stat_SG_1}, the latter is multiplied from the left by $\psi_{n'l'm'}^*(\vec{r})$ and integrated over $\vec{r}$. Once the occurring matrix elements between the Coulomb-Sturmian functions are evaluated \cite{zamastil_calculation_2007,schweiner_impact_2016}, the stationary Schr{\"o}dinger equation turns into a generalized eigenvalue problem
\begin{equation}
\label{eq:stat_SG}
\mathcal{A}(\gamma, f, b) \, \vec{v}_{\mathrm{stat}, i} = E_i\, \mathcal{B}\, \vec{v}_{\mathrm{stat}, i}
\ ,
\end{equation}
with the complex, symmetric matrix $\mathcal{A}$, the positive, semi-definite and real matrix $\mathcal{B}$, the eigenvector $\vec{v}$ and its corresponding generalized eigenvalue $E$. The matrix $\mathcal{A}$ is the matrix representation of $\mathcal{H}$ from \cref{eq:stat_SG_1} in the complete Coulomb-Sturmian basis, and the overlap matrix $\mathcal{B}$ occurs, since the Coulomb-Sturmian functions are not orthogonal with respect to the identity, but to the operator $1/r$. The generalized eigenvalue equation, obeying a complex symmetric form, now reveals resonance states with complex eigenvalues.
Due to the complex rotation the standard inner product has to be extended to the so-called $c$-product \cite{moiseyev_non-hermitian_2011}.

To solve the generalized eigenvalue problem~(\ref{eq:stat_SG}), for
simplicity the magnetic quantum number $m$ is set to zero, and the
Coulomb-Sturmian basis (cf.\ \cref{eq:cs_ansatz}) is cut off at a
principal quantum number $N_\mathrm{max}$. The finite-dimensional
eigenvalue problem with vector space size
$M=N_\mathrm{max}\left(N_\mathrm{max}+1\right)/2$ can be diagonalized
by the software package ARPACK \cite{lehoucq_arpack_1998}, which uses
the Implicitly Restarted Arnoldi Method (IRAM). For most of the
calculations we use $N_\mathrm{max}=35$, which corresponds to a vector
space size $M=630$.

\subsection{Time evolution of resonance occupation}

To study the temporal evolution of population for encirclings of EPs, the Schr{\"o}dinger dynamics has to be evaluated for a given loop $\gamma(t), f(t)$ in parameter space.
The time-dependent Schr{\"o}dinger equation
\begin{align}
\mathcal{B}\, \partial_t\vec{v} &=
-\i \mathcal{A}\left(\gamma(t), f(t)\right)\, \vec{v}
\label{eq:time_SG}
\end{align}
is a coupled first-order differential equation system, which is numerically solved by Cholesky factorization of the positive semi-definite matrix $\mathcal{B}$ and a Runge-Kutta integrator.
For the solution vector $\vec{v}(t)$ we use a spectral decomposition into the eigenvectors $\vec{v}_{i,  \mathrm{stat}}(t)$ of the actual Hamiltonian with the corresponding expansion coefficients $\alpha_i(t)$:
\begin{equation}
\vec{v}(t) = \sum_i \alpha_i(t) \vec{v}_{\mathrm{stat}, i}(t)
\ .
\end{equation}
In the literature this basis set is referred to as \emph{instantaneous basis}.
From the time-propagated vector $\vec{v}(t)$---which is the solution of \cref{eq:time_SG}---the $j$th expansion coefficient can be extracted via the projection
\begin{equation}
\alpha_j(t)=\vec{v}_{\mathrm{stat},j}^\mathrm{T}(t) \mathcal{B} \, \vec{v}(t)
\ .
\label{eq:alpha_projection}
\end{equation}
Note that here the $c$-product has to be taken into account.
The population within the $j$th resonance is given by the modulus squared of the expansion coefficient $\alpha_j$.

To gain numerically stable solutions of \cref{eq:time_SG} it is necessary to apply at regular time intervals projections of the vector $\vec{v}(t)$ onto the subspace of numerically converged resonances of the instantaneous Hamiltonian at time $t$. Otherwise couplings to non-converged resonances, which even can lie in the upper half of the complex energy plane, and thus would lead to an exponential growth during the time evolution, lead to instabilities in the population dynamics.

\subsection{\texorpdfstring{$\boldsymbol{2\times 2}$}{2x2} matrix  model}
\label{sec:matrix_model}
The calculation of the full quantum dynamics is numerically expensive. For example the computation time on a single core which is required to produce \cref{fig:pop_r_T} is approximately five days. Therefore we introduce an approximation method, which only describes the interaction between two commuting resonances and neglects the influence of side resonances. Besides the numerical aspect the approximation method provides insight into the influence of side resonances on population dynamics, when compared to the full calculation.

The approximation method is based on a $2\times 2$ Hamiltonian matrix $\mathcal{M}$, whose elements are expansions in the field strength parameters $\gamma$ and $f$ around the EP $\gamma_\mathrm{EP}, f_\mathrm{EP}$.
A priori the expansion coefficients are unknown.
Then we expand the barycentric coordinate
\begin{align}
\kappa(\gamma,f) &\equiv E_1 + E_2 = \mathop{\mathrm{tr}} \mathcal{M}
                   \nonumber \\
&= A+B(\gamma-\gamma_\mathrm{EP})+C(f-f_\mathrm{EP})
\label{eq:kappa}
\end{align}
of the eigenvalues $E_1, E_2$ in a polynomial of degree one and the relative coordinate
\begin{align}
\eta(\gamma,f) &\equiv (E_1-E_2)^2 = \mathop{\mathrm{tr}^2} \mathcal{M} - 4 \mathop{\mathrm{det}} \mathcal{M} \nonumber \\
&=  D+E(\gamma-\gamma_\mathrm{EP})+F(f-f_\mathrm{EP})+G(\gamma-\gamma_\mathrm{EP})^2 \nonumber \\
& \quad\, +H(f-f_\mathrm{EP})(\gamma-\gamma_\mathrm{EP})+I(f-f_\mathrm{EP})^2
\label{eq:eta}
\end{align}
in a polynomial of degree two with still a priori unknown coefficients $A,\dots , I$. The eigenvalues $E_1, E_2$ of $\mathcal{M}$ shall describe the parameter-dependent eigenvalues of the two commuting resonances obtained from the ``full'' problem described by  \cref{eq:stat_SG}. The coefficients $A,\dots, I$ are determined by solving \cref{eq:stat_SG} at nine points in parameter space $\gamma, f$ on an octagon. For more details see Ref.~\cite{feldmaier_rydberg_2016}.

To carry out dynamical calculations the explicit form of the $2\times 2$ Hamiltonian matrix $\mathcal{M}$ has to be known. Eqs.~(\ref{eq:kappa}) and (\ref{eq:eta}) fix the trace and determinant of $\mathcal{M}$. To uniquely define $\mathcal{M}$ two more constraints are required. As a third constraint we demand complex symmetry of $\mathcal{M}$, i.e. $\mathcal{M}=\mathcal{M}^\mathrm{T}$, since the same symmetry underlies the  full quantum mechanical eigenvalue \cref{eq:stat_SG}. With these three constraints the matrix $\mathcal{M}$ can be written in the form
\begingroup
\renewcommand*{\arraystretch}{1.5}
\begin{equation}
\label{eq:2x2_final}
\mathcal{M}(\gamma,f,c) = \begin{pmatrix}
\frac{\kappa}{2}+\frac{1}{4}\left(c+\frac{\eta}{c}\right)  &
\frac{\i}{4}\left( c-\frac{\eta}{c} \right) \\
\frac{\i}{4}\left( c-\frac{\eta}{c} \right)  &
\frac{\kappa}{2}-\frac{1}{4}\left(c+\frac{\eta}{c}\right)
\end{pmatrix}
\ .
\end{equation}
\endgroup
The derivation of \cref{eq:2x2_final} is given in \cref{app:2x2}. Due
to the lack of a fourth constraint on $\mathcal{M}$ a free complex parameter $c\neq 0$ emerges.
The time-dependent Schr{\"o}dinger equation for the approximation method reads
\begin{equation}
\label{eq:SG_2x2}
\i \partial_t \vec{v} = \mathcal{M}(\gamma(t),f(t),c) \vec{v}
\ .
\end{equation}
A natural question that arises is whether dynamical quantities calculated from \cref{eq:SG_2x2} depend on the choice of $c$. As shown in Ref.~\cite{burkhardt_matrixmodell_2016-1} a variation of the  free parameter $c$ over orders of magnitude has no impact on the population dynamics.

\section{Dynamical encircling exceptional points}
\label{sec:dyn_encircl}

Positions of some EPs within the two-dimensional parameter space
$\gamma, f$ have already been determined in
Refs.~\cite{Feldmaier2015,feldmaier_rydberg_2016}. For our calculations
we use a second-order EP at the parameter values \cite{Feldmaier2015}
\begin{equation}
f_\mathrm{EP}\equiv\num{3.176736E-4}\ \mathrm{a.u.},\ 
\gamma_\mathrm{EP}\equiv\num{1.445263E-2}
,
\end{equation}
or in SI units
\begin{equation}
F_\mathrm{EP}\equiv\SI{1.633870E8}{V/m},\ 
B_\mathrm{EP}\equiv\SI{3.396368E3}{T}
,
\end{equation}
which leads to coalescence at the eigenvalue
\begin{equation}
E_\mathrm{EP} = \num{-2.703665E-2} -\num{4.171979E-4}\,\i
\ .
\end{equation}

To reduce the dimensionality of the parameter loop around the EP we choose an elliptical parameterization for the field strength loop
\begin{subequations}
\begin{align}
f(t) &= f_\mathrm{EP}\left[ 1+r\cos\left( 2\pi\frac{t}{T}+\phi_0\right) \right] \ ,\\
\gamma(t) &= \gamma_\mathrm{EP}\left[ 1+r\sin\left( 2\pi\frac{t}{T}+\phi_0\right) \right]
\ ,
\end{align}
\label{eq:ellipse}
\end{subequations}
\begin{figure}
\centering
\includegraphics{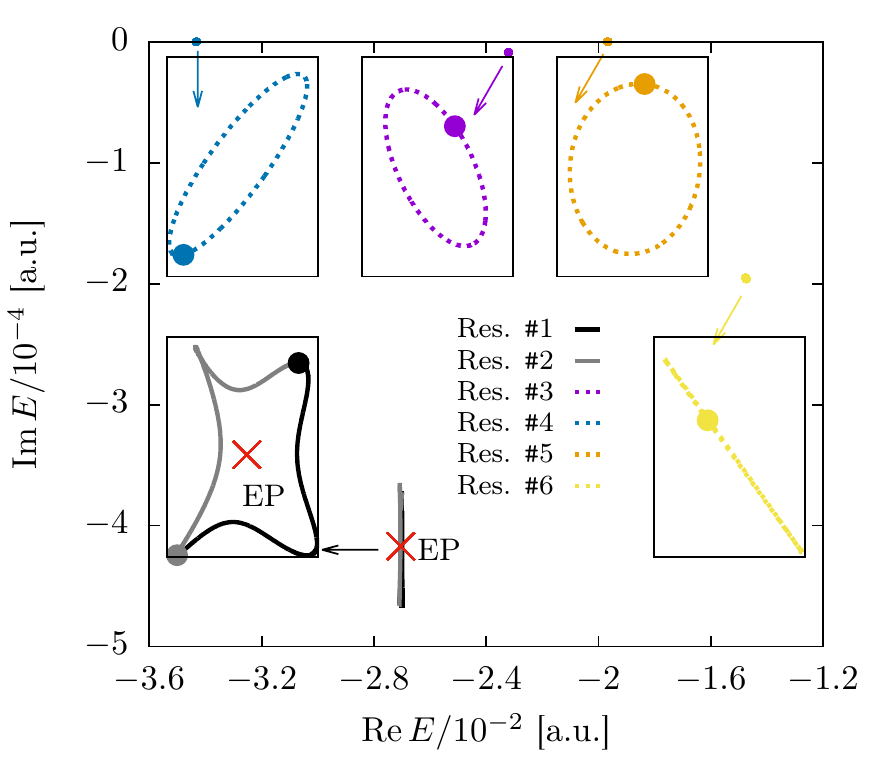}
\caption{Eigenvalue trajectories for an elliptical field strength loop according to \cref{eq:ellipse} with parameters $r=10^{-3}$ and $\phi_0=0$. The insets schematically show the trajectories of the isolated resonances. Start positions are marked with points. The EP resonances (\texttt{\#}1 and \texttt{\#}2) commute, whereas the neighboring four side resonances perform closed loops. Population is transferred from the eigenstate corresponding to start position of resonance \texttt{\#}1 to the eigenstate of start position at resonance \texttt{\#}2.
}
\label{fig:energies_r1e-3}
\end{figure}
containing three control parameters, namely the relative half-axis $r$, the starting angle $\phi_0$ on the ellipse and the loop duration $T$.

The population transfer protocol works as follows. At $t=0$ at the parameter values $f(0), \gamma(0)$ the whole system population is prepared in one of the two EP resonance states, which will be labeled by \texttt{\#1}, i.e. $|\alpha_1(t=0)|^2=1$, or equivalently $\vec{v}(t=0) = \vec{v}_{\mathrm{stat, 1}}(t=0)$. The dynamics of the resonance populations $|\alpha_i(t)|^2$ for a given elliptical parameter loop $f(t), \gamma(t)$, taking place in the time interval $t\in\left[0,T\right]$, are then calculated via \cref{eq:time_SG,eq:alpha_projection}. The question we ask is, which field strength loop $\gamma(t), f(t)$ maximizes the transferred population from the populated resonance \texttt{\#1} to the initially unpopulated EP resonance \texttt{\#2}. Since an adiabatic basis is used the resonances \texttt{\#1} and \texttt{\#2} commute for a parameter loop, and the transferred population is given by $|\alpha_1(t=T)|^2$. 

The three ellipse parameters $T,r$ and $\phi_0$ have great impact on the population dynamics. The encircling duration $T$ does not change the parameter path, but determines the adiabaticity of the parameter variation. With the ellipse radius $r$ the parameter path changes and the energy separation between resonances, and hence their mutual coupling strength, can be controlled. With $\phi_0$ the initial state can be shifted along the eigenvalue trajectory, and hence a specific path segment can be chosen. This allows us to adjust the decay rates of the EP resonances in a certain range.

In the following the influence of these three ellipse parameters on the population transfer shall be elaborated in detail to gain deeper insights into the transfer mechanism. First, we only vary the parameter $T$ or $r$, while the other two parameters are constant. Then $r$ and $T$ are varied simultaneously. Finally we study the impact of $\phi_0$ on the transfer.

\subsection{Influence of the encirling duration}

While the encircling duration $T$ is changing, the other ellipse parameters are constant at the values $r = 10^{-3}$ and $\phi_0=0$. Fig.~\ref{fig:energies_r1e-3} shows for this set of parameters the eigenvalue trajectories of the two EP resonances and the four neighboring side resonances for an elliptic encircling according to \cref{eq:ellipse}.
Since a small parameter radius $r$ is chosen the eigenvalue positions are only slightly modified. The insets schematically visualize, that side resonances (dotted lines) perform closed loops, while the EP resonances (solid lines) commute.

\begin{figure}
\centering
\includegraphics{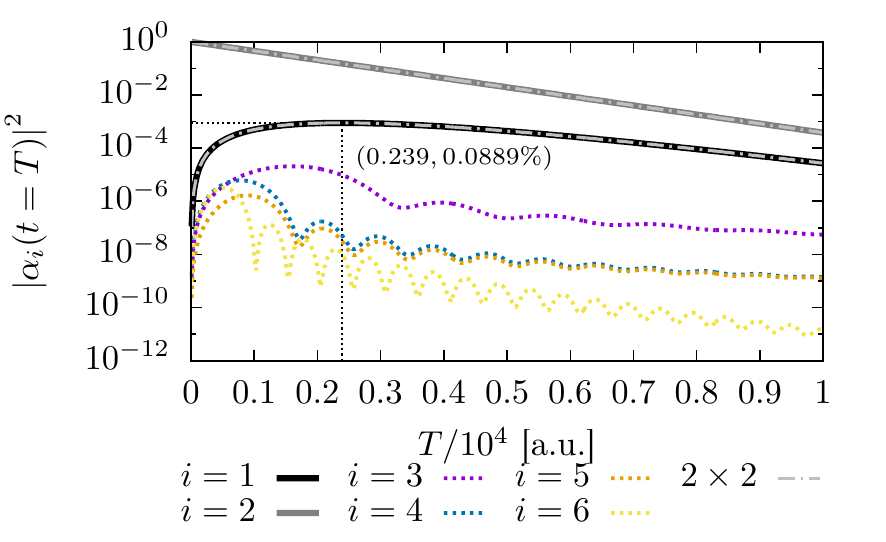}
\caption{Post-loop resonance populations $|\alpha_i(t=T)|^2$ for different encircling durations $T$ with $r=\num{1E-3}$ and $\phi_0=0$. The same colors as in \cref{fig:energies_r1e-3} are used for labeling the resonances. Since an adiabatic basis is used the states \texttt{\#1} and \texttt{\#2} commute after one parameter loop. The  maximum transferred population is $|\alpha_1(T_\mathrm{opt})|^2=\num{0.0889}{\%}$ at $T_\mathrm{opt}=\num{2.39E3}$. The dynamics of the EP resonances calculated with the $2\times 2$ matrix model (dash-dotted lines) coincide with the exact dynamics (solid lines).
}
\label{fig:pop_T}
\end{figure}

Since we are dealing with metastable, decaying states the timescale upon which the encircling takes place is a crucial parameter for optimizing the population transfer.
The  post-loop populations $|\alpha_i(t=T)|^2$ are shown for encircling durations $T\in\left(0,10^4\right]$ in \cref{fig:pop_T}.
For the EP resonances (solid lines) and side resonances (dotted lines) the same colors as in \cref{fig:energies_r1e-3} are used.
In general we find for the limiting case $\lim_{T\rightarrow 0}|\alpha_i(t=T)|=\delta_{2i}$ (with $\delta_{ij}$ the Kronecker delta), since the system cannot follow the fast fields and remains in its initial state, and for the limiting case $\lim_{T\rightarrow \infty}|\alpha_i(t=T)|=0$, since all resonances decay in time due to $\Im E_i < 0$.
Therefore the transferred population $|\alpha_1(T)|$ must go through a global maximum, which is located at $T=\num{2.39E3}$ with $0.0889\%$ population. Due to the small parameter radius ($r=10^{-3}$) the transfer is extremely small, which changes in the next subsection.

The dynamics according to the $2\times 2$ matrix model (dash-dotted lines) perfectly match the exact results.

\subsection{Influence of the parameter radius}

Now we focus on the population dynamics for different paths in parameter space around the EP. Therefore we vary the ellipse radius $r$, while the encircling duration $T=\num{2.5E3}$ and the starting angle $\phi_0=0$ are kept constant.

\begin{figure}
\centering
\includegraphics{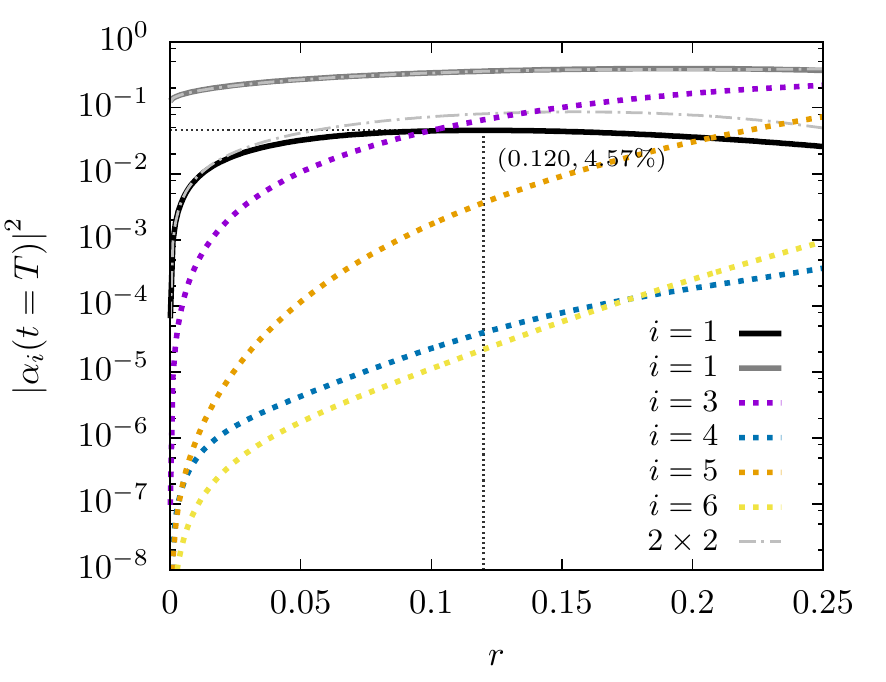}
\caption{Post-loop populations $|\alpha_i(t=T)|^2$ for different ellipse radii $r$, while $T=\num{2.5E3}$ and $\phi_0=0$. Since an adiabatic basis is used the population transfer is given by $|\alpha_1(T)|^2$. The latter strongly increases with $r$ for small $r$-values, goes through a global maximum at $r=0.12$ with $\num{4.57}{\%}$ transferred population, and decreases slightly for large $r$-values.
For large radii $r$ the approximate solution (dash-dotted lines)  shows deviations when compared to exact solution (solid lines).
}
\label{fig:pop_r}
\end{figure}

Fig.~\ref{fig:pop_r} shows the post-loop populations $|\alpha_i(T)|^2$ as a function of the relative ellipse radius $r$.
If the parameter loop is very close to the EP, the population transfer turns out to be small. In this region side resonance populations can be neglected. Optimal population transfer takes place at $r=0.12$ with $|\alpha_1|^2=4.57\%$. For a radius beyond this optimum point side resonances become stronger populated.

If one compares the EP resonance populations calculated by the exact
method (solid lines) and by the $2\times 2$ approximation method
(dash-dotted lines) it turns out that both methods agree well only for
loops close to the EP. For large values of $r$ deviations become
visible, due to two reasons. First, the $2\times 2$ matrix
$\mathcal{M}$ is a Taylor expansion around the EP (see Sec.~\ref{sec:matrix_model})
and therefore
only yields good results in a close vicinity of the EP. Second, the
approximation method only incorporates the EP resonances and neglects
side resonances, where the latter become strongly populated for large
$r$.

\begin{figure}
\centering
\includegraphics{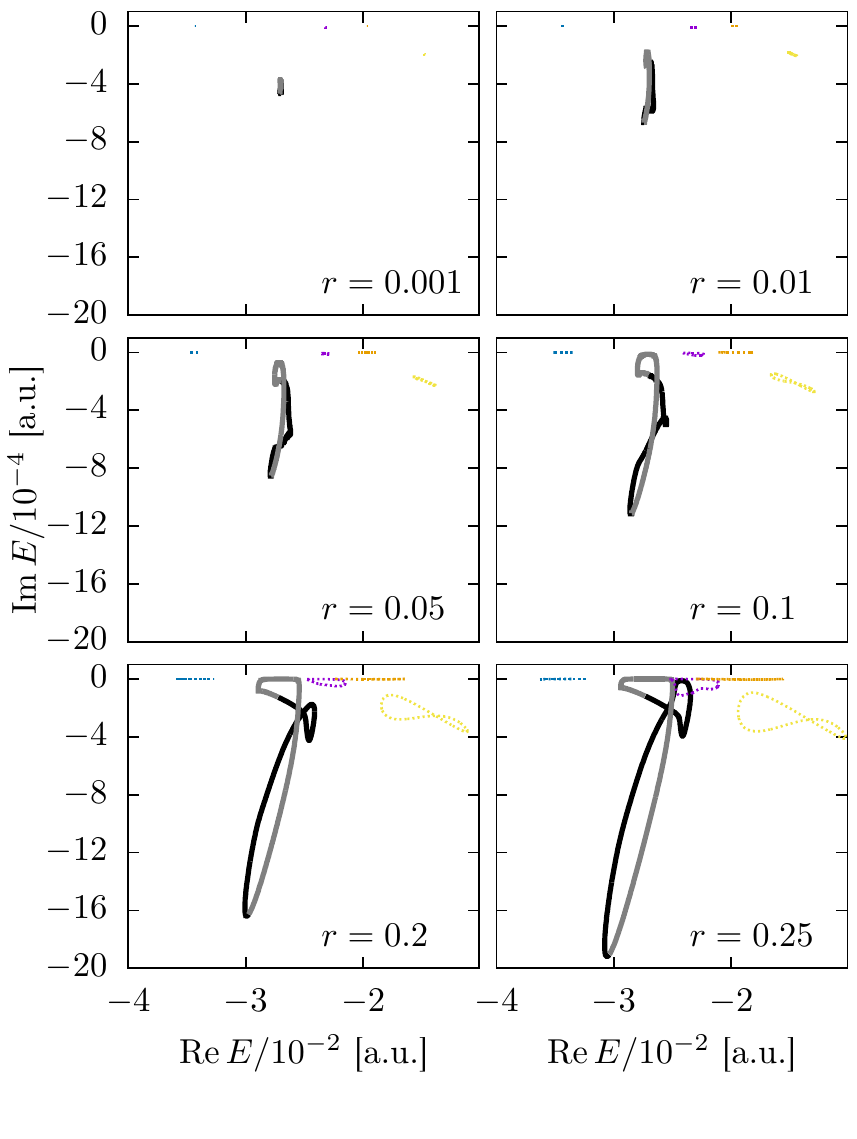}
\caption{Complex eigenvalue trajectories for elliptical parameter loops according to \cref{eq:ellipse} for six different ellipse radii $r$. For small $r$: EP resonances are strongly coupled. For large $r$: couplings to side resonances increase and the initially occupied resonance \texttt{\#}1 (black solid line) drifts deeper into the complex energy plane.
}
\label{fig:energies_r}
\end{figure}

To gain a better understanding of how the ellipse radius in parameter space affects the transfer it is helpful to regard the corresponding eigenvalue trajectories for a given parameter loop. In \cref{fig:energies_r} eigenvalue trajectories for six different values of $r$ are shown.
From left to right and top to bottom the radius parameter $r$ is increased, which causes the eigenvalue trajectories to carry out larger paths. The non-adiabatic coupling strength between two resonance states is inversely-proportional to their eigenvalue spacing. Hence, for small values of $r$ the strong coupling between both EP resonances hampers population transfer. With growing parameter radius $r$ the coupling of the initially populated resonance to the other EP resonance diminishes, but at the same time couplings to neighboring resonances grow, which manifests itself in increasing side resonance populations (see \cref{fig:pop_r}). However, the coupling of the initially populated resonance to side resonances cannot be the only reason for the decreasing population transfer in the large radius regime, since the  dynamics according to the $2\times 2$ model reveals the same behavior, although there are no couplings to side resonances implemented. The second reason for falling transfer with increasing $r$ is the circumstance that the energy trajectory of the resonance \texttt{\#}1 drifts deeper into the complex energy plane for growing $r$ and hence decays more rapidly in time.


\subsection{Optimal encircling duration and parameter radius}

\begin{figure}
\centering
\includegraphics{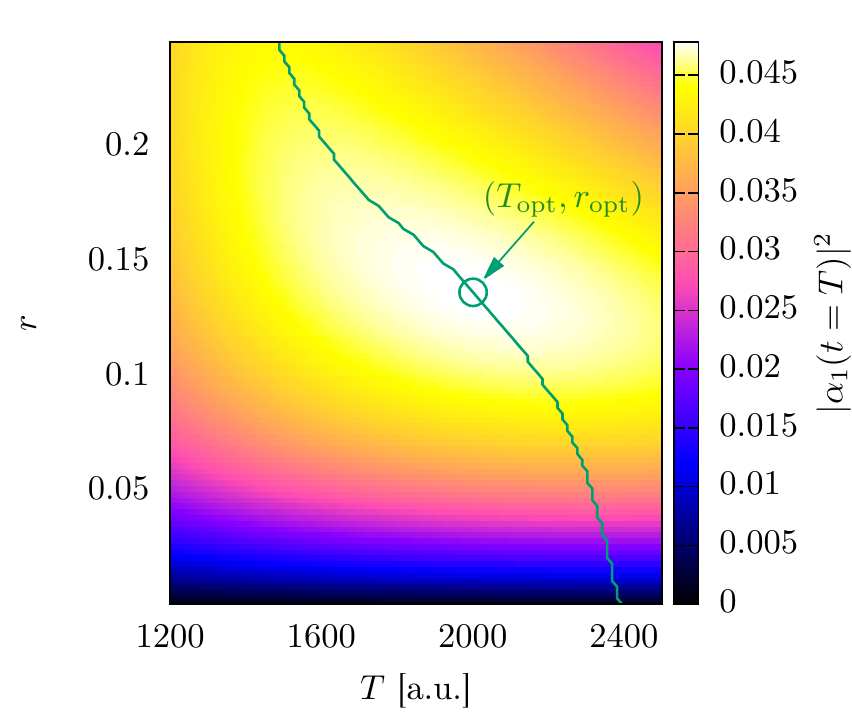}
\caption{Population transfer $|\alpha_1(T)|^2$ as function of the ellipse radius $r$ and encircling duration $T$. The green curve is the optimal encircling duration for given $r$. The optimal encircling duration decreases with increasing $r$, since the decay rate of resonance \texttt{\#}1 increases (cf. \cref{fig:energies_r}). For the global optimum $T_\mathrm{opt}=\num{2.001E3}$, $r_\mathrm{opt}=\num{0.1368}$ the transfer amounts to $4.783\%$.
}
\label{fig:pop_r_T}
\end{figure}

To determine the optimal loop parameters $T_\mathrm{opt}$ and $r_\mathrm{opt}$ for which the population transfer $|\alpha_1(T)|^2$ becomes extremal,  the transfer $|\alpha_1(T)|^2$ was calculated on a $100\times 100$ grid in $r$-$T$-space (see \cref{fig:pop_r_T}).

The green curve refers the optimal encircling duration $T$ to each radius parameter $r$. It is remarkable that for increasing ellipse radius $r$ the optimal encircling duration becomes shorter. From \cref{fig:energies_r} it follows that for increasing $r$ the EP eigenvalue trajectories extend deeper into the negative complex plane. Therefore, in terms of population transfer, a long residence within strongly decaying states gets penalized.

Along the green curve there exists a global optimum point
$(T_\mathrm{opt}=\num{2.001E3},\ r_\mathrm{opt}=\num{0.1368})$, 
for which $\partial_T|\alpha_1|^2\big|_{(r_\mathrm{opt},T_\mathrm{opt})} = \partial_r|\alpha_1|^2\big|_{(r_\mathrm{opt},T_\mathrm{opt})} =0$. The value of transferred population there amounts to $|\alpha_1(T)|^2\big|_{(r_\mathrm{opt},T_\mathrm{opt})} = 4.783\%$.

\subsection{Influence of the starting angle}

Finally the impact of the ellipse starting angle $\phi_0$ from \cref{eq:ellipse} on the transfer shall be elaborated. It is an important control parameter, since it determines the path segments the EP resonances will cover during one revolution in parameter space.
In particular the global decay of the adiabatic evolution of the $i$th resonance eigenstate
\begin{equation}
\left| \psi_i\right|^2 \sim \exp{\left[ \int_{t=0}^T 2 \Im E_i(t) \, \mathrm{d}t \right]}
\end{equation}
is affected by $\phi_0$.

\begin{figure}
\centering
\includegraphics[scale=0.98]{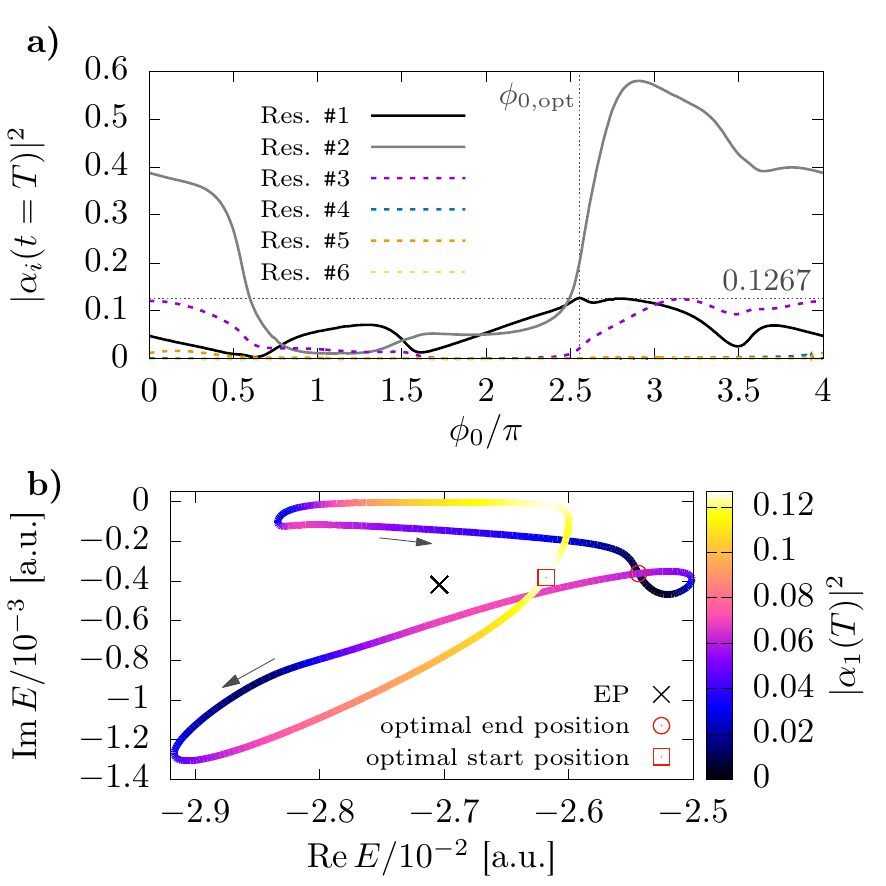}
\caption{%
(a) Post-loop populations $|\alpha_i(T)|^2$ as a function of the starting angle $\phi_0$ of the parameter ellipse. The $4\pi$-periodic transfer reaches the global maximum $|\alpha_1(T)|^2=12.67\%$ at $\phi_{0,\mathrm{opt}}=2.55276\pi$.
In (b) the transfer $|\alpha_1(T)|^2$ (color bar) is related to the corresponding start position of the initially populated resonance \texttt{\#}1 in energy space. The arrows indicate the direction of rotation of the eigenvalue for growing $t$ in \cref{eq:ellipse}. Transfer is maximal for the eigenvalue path of resonance \texttt{\#}1, which exhibits the smallest global decay.
}
\label{fig:phi0}
\end{figure}

The ellipse radius $r$ and encircling duration $T$ are held fixed at the previously optimized values $r_\mathrm{opt},\ T_\mathrm{opt}$. In \cref{fig:phi0}(a) the post-loop resonance populations $|\alpha_i(T)|^2$ are displayed as a function of the starting angle $\phi_0$.
$\phi_0$ goes over $4\pi$, since we wish to consider all starting positions of the initially populated resonance \texttt{\#}1 in energy space, which has a periodicity of $4\pi$. By variation of $\phi_0$ the transfer can be considerably optimized. At $\phi_{0,\mathrm{opt}}=2.55276\pi$ a transfer of $12.67\%$ is reached. Of course the transfer for an elliptical parameterization depends on all three parameters $r$, $T$ and $\phi_0$. To find the global maximum of the transfer all three parameters would have to be varied simultaneously, which goes beyond the scope of this paper.

To relate the $\phi_0$-dependent transfer to the eigenvalue trajectories, \cref{fig:phi0}(b) shows the start positions of the initially populated resonance \texttt{\#}1 and the corresponding transfer $|\alpha_1(T)|^2$ (color bar). The start (end) position for which the transfer is maximal is marked. If the direction of rotation is taken into account, it turns out that the ``best'' path minimizes the global decay of resonance \texttt{\#}1. On the other hand, for the starting angle $\phi_0 = \phi_{0,\mathrm{opt}}-2\pi = 0.55276\pi$ the initial population is prepared in the resonance marked by the rectangle in \cref{fig:phi0}(b). For this situation the populated resonance goes the most dissipative path and the transfer obeys a global minimum. 

\begin{figure}
\centering
\includegraphics[scale=0.98]{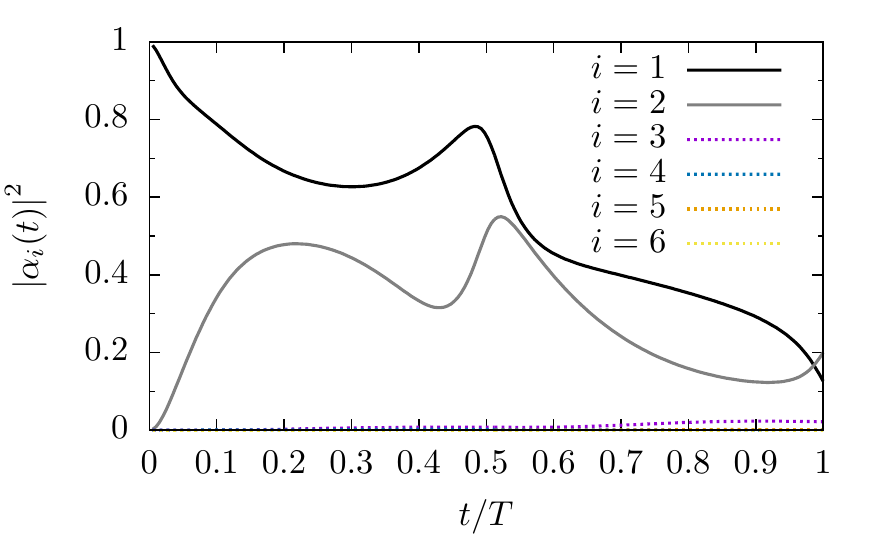}
\caption{%
Population dynamics $|\alpha_i(t)|^2$ for the optimal loop parameters $T_\mathrm{opt}=\num{2.001E3}$, $r_\mathrm{opt}=\num{0.1368}$ and $\phi_{0,\mathrm{opt}}= 2.55276\pi$. The amount of transferred population denotes $12.67\%$.}
\label{fig:dynamics_optimum_point}
\end{figure}

The explicit population dynamics of the resonances for the optimal parameter loop with parameters $T_\mathrm{opt}=\num{2.001E3}$, $r_\mathrm{opt}=\num{0.1368}$ and $\phi_{0,\mathrm{opt}}= 2.55276\pi$ is given in \cref{fig:dynamics_optimum_point}.

\section{Conclusion}
\label{sec:conclusion}

We studied the time evolution of population for resonance states within the hydrogen atom in parallel electric and magnetic fields. In order to transfer population from one resonance to another the commutation behavior of eigenstates for closed parameter loops around EPs is used.

The presented $2\times 2$ matrix model reduces the high-dimensional vector space and is therefore numerically very efficient. It yields good results for the population dynamics (compared with the exact calculations) in a close vicinity of the EP. Far away from the EP deviations occur, since (i) the $2\times 2$ matrix is based on a Taylor expansion around the EP, and (ii) couplings to side resonances, which are not included in the model, become stronger.

We have shown, that for a given parameter loop an optimal encircling duration exists which maximizes the transfer. The eigenvalue trajectories of the side resonances and EP resonances can be exploited to determine the optimal shape of the parameter loop. Since the non-adiabatic coupling strength between two resonances is inversely-proportional to their energy spacing, the eigenvalue of the initially populated state should be well-separated from all the other resonances along its trajectory to avoid population currents to them. Furthermore the populated resonance should exhibit a small global decay, which can be realized if the absolute value of the imaginary part of the eigenvalue is small along the trajectory.

The EP investigated here corresponds to a magnetic field strenth of
approximately three thousand teslas, which is inaccessible to present
experiments. In principle EPs should also exist at laboratory field
strengths. From the numerical perspective these EPs are however hard
to calculate, since the required basis size strongly grows with
decreasing field strengths. Cuprous oxide should allow for the
observation of EPs at laboratory field strengths and low principal
quantum numbers, however, in that case the complete valence band
structure of Cu$_2$O must be taken into account for detailed
comparisons between experiment and theory, see e.g. \cite{Schweiner2017}.

\appendix

\section{Hamiltonian of the \texorpdfstring{$\boldsymbol{2\times 2}$}{2x2} matrix model}
\label{app:2x2}

The behaviour of the eigenvalues of the two EP resonances with respect to the external field parameters $\gamma$ and $f$ is described by $\kappa(\gamma,f)$ and $\eta(\gamma,f)$ in \cref{eq:kappa,eq:eta}. Here we derive an expression for the $2\times 2$ matrix Hamiltonian $\mathcal{M}$, which is needed for time-dependent calculations. In the quantities $\kappa$ and $\eta$ and by assuming that $\mathcal{M}$ is complex symmetric, the Hamiltonian matrix takes the general form
\begin{equation}
\mathcal{M}=
\begin{pmatrix}
g_1(\kappa,\eta)  & f_2(\kappa, \eta) \\
f_2(\kappa, \eta) & g_2(\kappa,\eta)
\end{pmatrix}
\ .
\end{equation}
By applying the two constraints
\begin{equation}
\kappa = \mathop{\mathrm{tr}}\mathcal{M} = g_1+g_2
\label{eq:ap1_kappa}
\end{equation}
and
\begin{equation}
\eta = \mathop{\mathrm{tr}^2}\mathcal{M} -4\mathop{\mathrm{det}}\mathcal{M}
\overset{(\ref{eq:ap1_kappa})}{=}
4\left( \frac{\kappa}{2}-g_1 \right)^2 + 4f_2^2
\label{eq:app_2x2_eta}
\end{equation}
on the general form of $\mathcal{M}$, it follows that $g_1$ and $g_2$ take the form
\begin{equation}
g_{1,2}(\kappa) \equiv \frac{\kappa}{2} \pm f_1(\eta)
\end{equation}
and the function $f_1$ is independent of $\kappa$, since the rhs. of \cref{eq:app_2x2_eta} has to be independent of $\kappa$. This leads to the intermediate form
\begin{equation}
\mathcal{M}=
\begin{pmatrix}
\frac{\kappa}{2} +f_1(\eta)  &  f_2(\eta) \\
f_2(\eta)  & \frac{\kappa}{2} -f_1(\eta)
\end{pmatrix}
\end{equation}
of the Hamiltonian.
As an ansatz for the functions $f_1$ and $f_2$ we chose
\begin{equation}
f_i(\eta) = a_i+b_i\eta
\ .
\end{equation}
Now the condition
\begin{equation}
\eta = \mathop{\mathrm{tr}^2}\mathcal{M} -4\mathop{\mathrm{det}}\mathcal{M}
= 4(f_1^2+f_2^2)
\end{equation}
yields by comparing the coefficients
\begin{equation}
b_1^2+b_2^2=a_1^2+a_2^2=0
\quad\mathrm{and}\quad
a_1b_1+a_2b_2=\frac{1}{8}
\ .
\end{equation}
A solution to this equation system is
\begin{equation}
a_2=\mathrm{i}a_1
\ ,\quad
b_2=-\mathrm{i}b_1
\ ,\quad
b_1=\frac{1}{16a_1}
\ .
\end{equation}
By defining the free parameter $c\equiv 4a_1$ the functions $f_i(\eta)$ are
\begingroup
\renewcommand*{\arraystretch}{1.5}
\begin{equation}
f_1=\frac{1}{4}\left( c+\frac{\eta}{c}\right)
\ , \quad
f_2=\frac{\mathrm{i}}{4}\left( c-\frac{\eta}{c}\right)
\end{equation}
and the final form of the matrix Hamiltonian is
\begin{equation}
\mathcal{M}=
\begin{pmatrix}
\frac{\kappa}{2} + \frac{1}{4}\left( c+\frac{\eta}{c} \right) & \frac{\mathrm{i}}{4}\left( c-\frac{\eta}{c} \right) \\
 \frac{\mathrm{i}}{4}\left( c-\frac{\eta}{c} \right) & \frac{\kappa}{2} - \frac{1}{4}\left( c+\frac{\eta}{c} \right)
\end{pmatrix}
\ .
\end{equation}
\endgroup

%

\end{document}